# Thermal Conductance across $\beta$-Ga$_2$O$_3$-diamond Van der Waals Heterogeneous Interfaces

Zhe Cheng[1], Luke Yates[1], Jingjing Shi[1], Marko J. Tadjer[2], Karl D. Hobart[2], Samuel Graham[1, 3, *]

[1] George W. Woodruff School of Mechanical Engineering, Georgia Institute of Technology, Atlanta, Georgia 30332, USA

[2] U.S. Naval Research Laboratory, 4555 Overlook Ave SW, Washington, DC 20375, USA

[3] School of Materials Science and Engineering, Georgia Institute of Technology, Atlanta, Georgia 30332, USA

[*] Corresponding author: sgraham@gatech.edu

KEYWORDS: thermal conductance, $\beta$-gallium oxide, Van der Waals interfaces, Time-domain Thermoreflectance, Landauer approach

**ABSTRACT**

Because of its ultra-wide bandgap, high breakdown electric field, and large-area affordable substrates grown from the melt, β-$Ga_2O_3$ has attracted great attention recently for potential applications of power electronics. However, its thermal conductivity is significantly lower than those of other wide bandgap semiconductors, such as AlN, SiC, GaN, and diamond. To ensure reliable operation with minimal self-heating at high power, proper thermal management is even more essential for $Ga_2O_3$ devices. Similarly to the past approaches aiming to alleviate self-heating in GaN high electron mobility transistors (HEMTs), a possible solution has been to integrate thin $Ga_2O_3$ membranes with diamond to fabricate $Ga_2O_3$-on-diamond lateral metal-semiconductor field-effect transistor (MESFET) or metal-oxide-semiconductor field-effect transistor (MOSFET) devices by taking advantage of the ultra-high thermal conductivity of diamond. Even though the thermal boundary conductance (TBC) between wide bandgap semiconductor devices such as GaN HEMTs and a diamond substrate is of primary importance for heat dissipation in these devices, fundamental understanding of the $Ga_2O_3$/diamond thermal interface is still missing. In this work, we study the thermal transport across the interfaces of $Ga_2O_3$ exfoliated onto a single crystal diamond. The Van der Waals bonded $Ga_2O_3$-diamond TBC is measured to be 17 -1.7/+2.0 MW/m$^2$-K, which is comparable to the TBC of several physical-vapor-deposited metals on diamond. A Landauer approach is used to help understand phonon transport across perfect $Ga_2O_3$-diamond interface, which in turn sheds light on the possible TBC one could achieve with an optimized interface. A reduced thermal conductivity of the $Ga_2O_3$ nano-membrane is also observed due to additional phonon-membrane boundary scattering. The impact of the $Ga_2O_3$–substrate TBC and substrate thermal conductivity on the thermal performance of a power device are modeled and discussed. Without loss of generality, this study is not only important for $Ga_2O_3$ power electronics applications which would not be realistic without a thermal management solution, but also for the fundamental thermal science of heat transport across Van der Waals bonded interfaces.

## 1. INTRODUCTION

As an emerging ultra-wide bandgap semiconductor material, β-Ga$_2$O$_3$ has shown favorable properties for use in power electronics applications, such as an ultra-wide bandgap (4.8 eV) and high critical electric field (8 MV/cm), which predict a Baliga figure of merit that is 3214 times that of Si.[1] However, the thermal conductivity of bulk β-Ga$_2$O$_3$ (10-30 W/m-K, depending on crystal orientation) is at least one order of magnitude lower than those of other wide bandgap semiconductors, for instance, GaN (230 W/m-K), 4H-SiC (490 W/m-K), and diamond (>2000 W/m-K).[2-3] To utilize Ga$_2$O$_3$ in high frequency and high power switching applications, proper thermal management is essential to avoid device degradation due to poor thermal reliability. This will require the use of high thermal conductivity pathways to pull the heat out of the Ga$_2$O$_3$ devices efficiently through interfacial contacts with low thermal boundary resistance. With its ultra-high thermal conductivity, diamond, which has been extensively studied to dissipate localized self-heating from electronics such as AlGaN/GaN HEMTs, is a possible solution for Ga$_2$O$_3$ devices as well.[4-7]

Recently, mechanically exfoliated Ga$_2$O$_3$ nano-membranes have been utilized to fabricate high-current transistors.[8-13] A record high drain current has been achieved in an exfoliated Ga$_2$O$_3$ field-effect transistor with diamond substrates.[14] This work demonstrates good device performance but has not quantified the heat transport across the Ga$_2$O$_3$-diamond interface as the Ga$_2$O$_3$ nano-membranes were adhered to diamond via Van der Waals forces. The TBC of mechanically joined materials could be as low as 0.1 MW/m$^2$-K while the interfacial thermal conductance of transfer-printed metal films is in the range of 10-40 MW/m$^2$-K.[15-19] Thermal transport across Van der Waals interfaces is limited by the real contact area and low phonon transmission due to weak adhesion energy even if there exists the possibility to achieve a high TBC.[20-22] Thermal transport across these interfaces remains an open issue due to the limited amount of experimental data available in the literature. Therefore, it is of great significance to study the thermal conductance across $Ga_2O_3$-diamond interfaces for both real-world power electronics applications and fundamental thermal science of heat transport across Van der Waals interfaces.

In this work, we have mechanically exfoliated a (100) oriented $Ga_2O_3$ nano-membrane from an EFG-grown commercial (-201) $Ga_2O_3$ substrate (Novel Crystal Technology, Japan) with medium-tack dicing saw tape and transferred it on a single crystal (100) CVD diamond substrate (Element Six).[23] Time-domain thermoreflectance (TDTR) was used to measure the $Ga_2O_3$-diamond TBC and $Ga_2O_3$ thermal conductivity. By combining picosecond acoustic technique and the membrane thickness measured by an Atomic Force Microscopy (AFM), the phonon group velocity across the membrane thickness direction is obtained. Moreover, we use a Landauer approach to calculate phonon transport across $Ga_2O_3$-diamond interfaces. Additionally, the effects of $Ga_2O_3$-substrate TBC and substrate thermal conductivity on thermal performance of a power electronics are modelled by an analytical solution.

## 2. RESULTS AND DISCUSSION

Figure 1(a) shows a part of the sample scanned by AFM. A blanket layer of Al (~80 nm) was deposited to serve as the TDTR transducer. As the size of the $Ga_2O_3$ nano-membrane was approximately 11 μm x 70 μm, a CCD camera integrated in the TDTR system was used to help locate the sample. To obtain the thermal conductivity of the diamond substrate, TDTR was performed on the area which was not covered by the Ga$_2$O$_3$ nano-membrane. The thermal conductivity of the single crystal diamond substrate was determined to be 2169 ±130 W/m-K, which is very close to other values reported in the literature.[24] This value was used in the data analysis and parameter fittings from the TDTR measurements on the Ga$_2$O$_3$ bonded to diamond. The Al-diamond TBC is also determined in the measurement to be 34 MW/m$^2$-K. Figure 1(b) shows the picosecond acoustic echoes obtained during TDTR measurements. The observed echoes correspond to strain waves that are reflected at interfaces.[25] Figure 1(c) shows the strain wave traveling distance for each echo. For echo 1, a small valley shows up before a peak, indicating the loose bonding of Van der Waals forces at the interface. In this scenario, we pick the middle point of the valley and peak as the echo point ($t_1$=25 ps). The sound speed of Al is 6420 m/s[25] so the Al thickness was determined to be 80 nm ($d_1 = v_{Al} * t_1/2$). For echo 2 (154 ps=25 ps+129 ps), it relates to the Ga$_2$O$_3$-diamond interface. The travelling time in Ga$_2$O$_3$ is 129 ps. The thickness of the Ga$_2$O$_3$ layer was determined as 427±3 nm by an AFM. Then the longitudinal phonon group velocity of Ga$_2$O$_3$ in the direction perpendicular to (100) plane is determined as 6620 m/s, which matched very well with DFT-calculated value (6809 m/s) we will discuss more later. Echo 3 is the strain wave bouncing back from the Al-air interface after coming back from the Ga$_2$O$_3$-diamond interface. The strain wave travels across Al layer again and bounces back from the Al-Ga$_2$O$_3$ interface. The traveling time equals 179 ps (=25 ps+129 ps+25 ps).

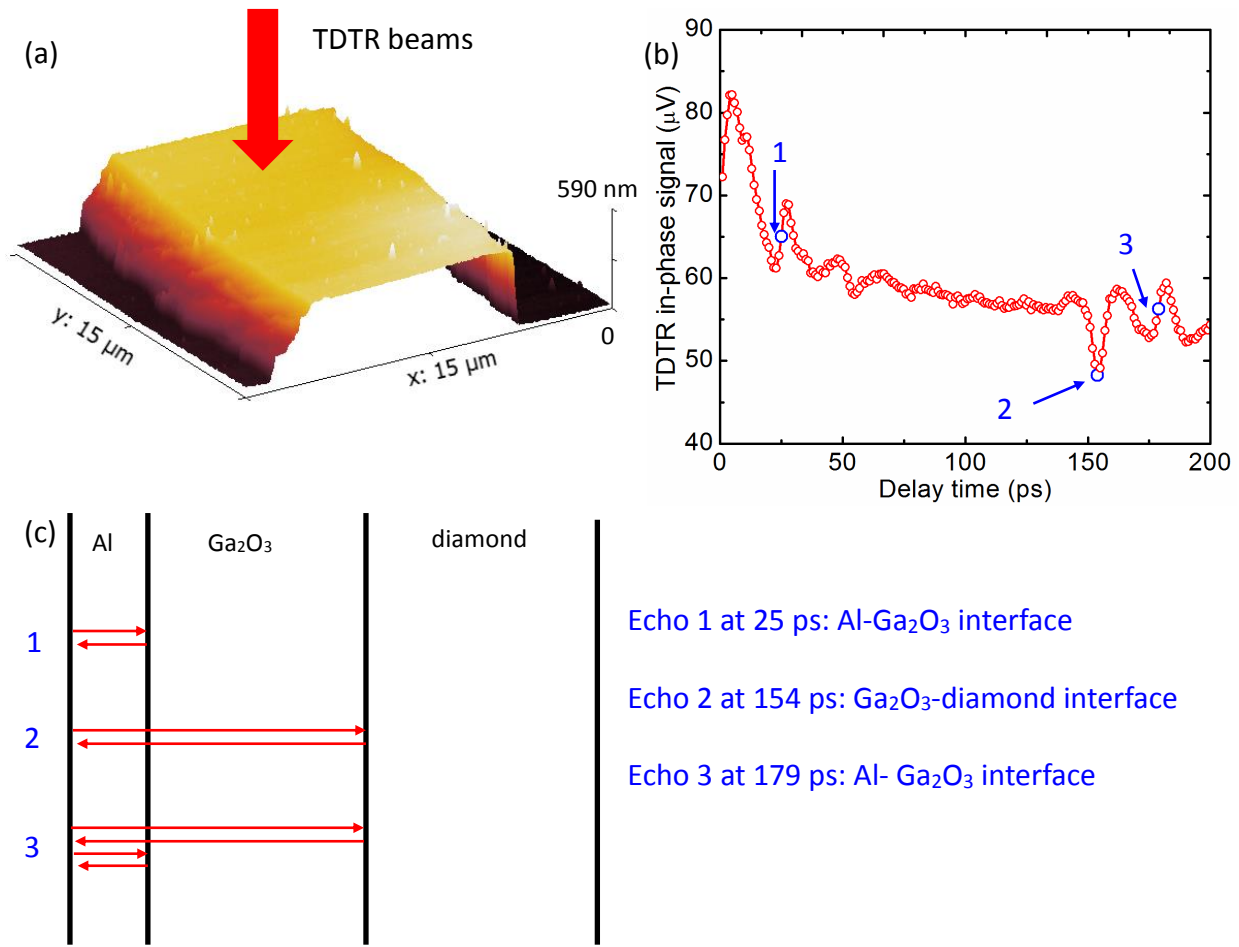

Figure 1. (a) AFM image of the center of the Ga$_2$O$_3$ sample. (b) Picosecond acoustic echoes obtained in the TDTR measurements. (c) Echoes which relate to strain wave bouncing back at interfaces.

The surface roughness of the two surfaces affects the real contact area, and correspondingly, the thermal conductance across this interface. To measure the surface roughness, AFM was used to scan the diamond substrate surface and the top surface of the Ga$_2$O$_3$ membrane. The bottom surface of the Ga$_2$O$_3$ membrane should be similar to or smoother than the top surface.[26-27] The surface images are shown in Figure 2(a-b). The RMS roughness of diamond and Ga$_2$O$_3$ surfaces are 3.39 ±0.91 nm and 3.23±0.93 nm, respectively. The sensitivity of our TDTR measurements was determined by considering a fractional change in the thermoreflectance signal due to a fractional change in the independent parameters.[4] Figure 2(c) shows the TDTR sensitivity of the TBC of the $Ga_2O_3$-diamond interface, thermal conductivity of the $Ga_2O_3$ membrane, and the TBC of the Al-$Ga_2O_3$ interface with a modulation frequency of 2.2 MHz and a 20 X objective (pump radius 4.9 μm and probe radius 3.0 μm). The sensitivity of the TBC of the $Ga_2O_3$-diamond interface is very large, resulting in accurate measurements of this parameter. Figure 2(d) shows good agreement between the experimental data and the fitted curve.

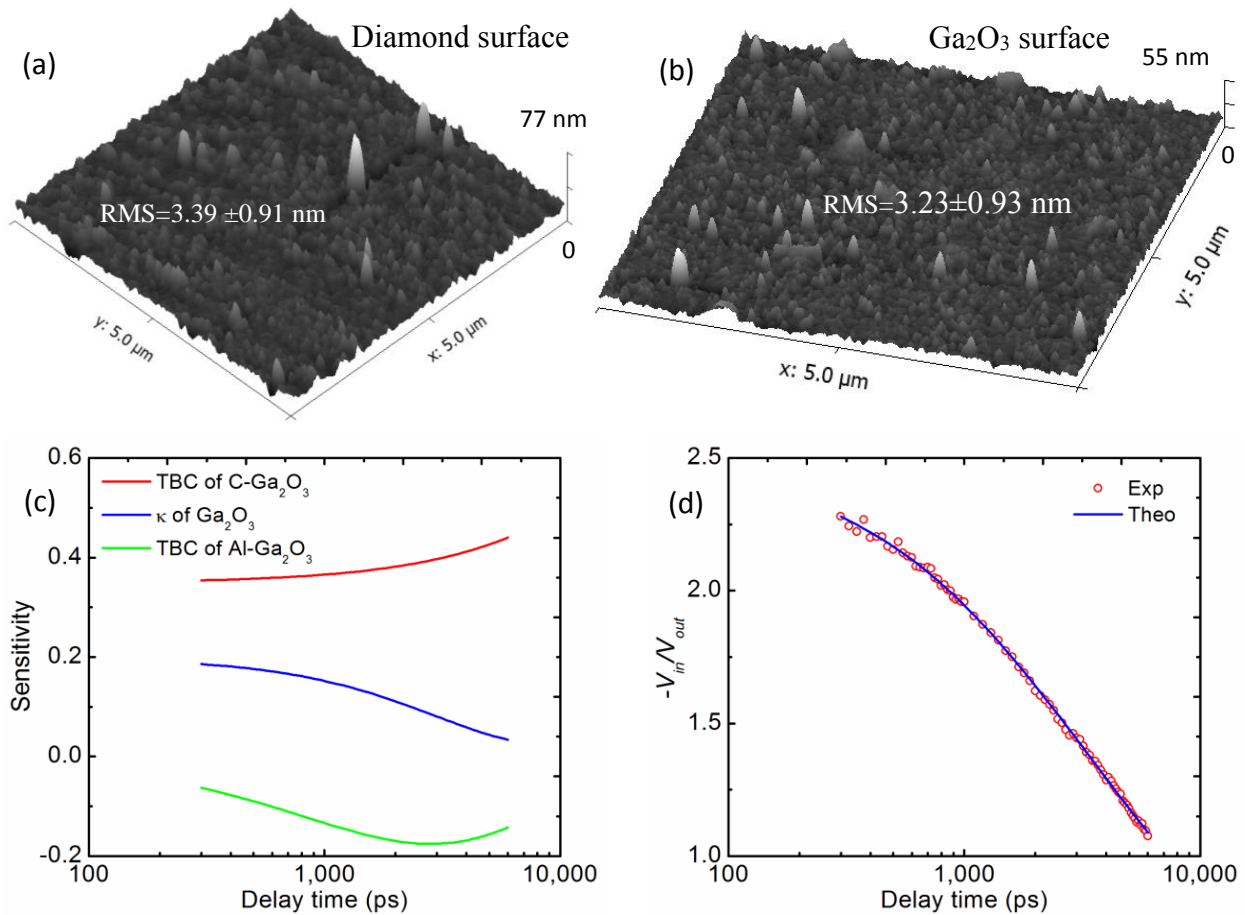

Figure 2. (a) AFM scanned surface roughness of the diamond substrate. (b) AFM scanned surface roughness of the top surface of the $Ga_2O_3$ nano-membrane. (c) TDTR sensitivity of TBC of the

Ga$_2$O$_3$-diamond interface, thermal conductivity of Ga$_2$O$_3$, and the TBC of Al- Ga$_2$O$_3$ interface with a modulation frequency of 2.2 MHz and a 20 X objective. (d) TDTR data fitting with modulation frequency of 2.2 MHz and 20 X objective.

The TBC of the Ga$_2$O$_3$-diamond interface was measured to be 17 -1.7/+2.0 MW/m$^2$-K and is compared with the TBC of several other diamond interfaces and transferred interfaces in Table 1. The error bars of TDTR measurements is estimated by a Monte Carlo method.[28] The Ga$_2$O$_3$-diamond TBC is in the TBC range of the transferred metal films on silicon, SiO$_2$, and sapphire substrates and is comparable to the TBC of physical-vapor-deposited metals on diamond. The Ga$_2$O$_3$-diamond TBC is mainly affected by three factors: the weak Van der Waals force between Ga$_2$O$_3$ and diamond, the small contact area at the interface, and the large phonon density of states (DOS) mismatch between Ga$_2$O$_3$ and diamond. For the Ga$_2$O$_3$-diamond interface, the two materials are bonded by the Van der Waals force. Interfacial bonding affects thermal conductance significantly.[29] For instance, the TBC of covalent bonded interfaces is much larger than that of the Van der Waals force bonded interfaces because phonon transmission is very low due to the weak adhesion energy of Van der Waals interfaces.[21, 29-30] Moreover, diamond is non-polar, so no dipolar-dipolar attraction exists at the interface. The Van der Waals force at the Ga$_2$O$_3$-diamond interface should be weaker than those of other polar material interfaces. This further decreases the TBC. In terms of contact area at the interface, the fractional areal coverage of transferred metal films on silicon or sapphire substrates could reach 25% because of plastic deformation and capillary forces.[20] The transferred metal thin films (Au) are very soft so the contact area between the metals and substrates could be very large under pressure during the transfer process. However, for the Ga$_2$O$_3$-diamond interface, diamond is known as one of the hardest materials and $Ga_2O_3$ is much harder than Au. We can see very small pillars forming surface roughness of both the $Ga_2O_3$ and diamond surfaces according to the AFM images. We speculate these pillars may enlarge the contact area at the interface and enhances thermal transport. Our measured $Ga_2O_3$-diamond TBC is comparable to the TBC of physical-vapor-deposited metals on diamond and reaches more than one fourth of the TBC of the chemical-vapor-deposited diamond on silicon even though the contact area of these deposited interfaces are much larger than that of $Ga_2O_3$-diamond interface.

Here, we define "diamond interface" as any interface one side of which is diamond. Diamond has ultra-high thermal conductivity due to the light carbon atom and strong covalent bond among carbon atom, which leads to an ultra-high Debye frequency and cutoff frequency. As a result, the phonon DOS match of diamond and other materials are very poor and the TBC of diamond interfaces are very low. When integrating diamond with other materials to take advantage of its high thermal conductivity, the low TBC of diamond interfaces is usually the bottleneck. This highlights the motivation to study thermal transport across diamond interfaces for both fundamental science and real-world applications. By taking all these into consideration, we could conclude that the measured $Ga_2O_3$-diamond TBC is relatively quite high. On one hand, this relatively high TBC helps to explain why a $Ga_2O_3$ field-effect transistor observed record-high drain current on diamond.[14] On the other hand, it shows that thermal transport across Van der Waals interfaces are relatively good from a fundamental viewpoint.

Table 1. TBC of several diamond interfaces and transferred interfaces

| Interfaces | TBC (MW/m$^2$-K) | Fabrication Conditions |
| --- | --- | --- |

| | | | |
|---|---|---|---|
| Our work | Ga$_2$O$_3$-diamond | 17 | Transferred van der Waals interfaces |
| Ref.[17] | Au-Si/SiO$_2$/Al$_2$O$_3$ | 10-40 | Transferred van der Waals interfaces |
| Ref.[31] | Bi-H-diamond | 8 | Physical vapor deposition |
| Ref.[31] | Pb-diamond | 19 | Physical vapor deposition |
| Ref.[31] | Pb-H-diamond | 15 | Physical vapor deposition |
| Ref.[7] | Si-diamond | 63 | Chemical vapor deposition |

The thermal conductivity of the Ga$_2$O$_3$ nano-membrane was measured as 8.4 ±1.0 W/m-K, which is lower than the value of bulk Ga$_2$O$_3$ in this direction (about 13 W/m-K).[3] The thickness dependent thermal conductivity of Ga$_2$O$_3$ thin films from literature and this work are summarized in Figure 3. The thermal conductivity Ga$_2$O$_3$ thin films decreases with film thickness. The phonon mean free path in bulk Ga$_2$O$_3$ ranges from several nm to several um.[32] Phonons with long mean free path scatter with film boundaries. The additional film boundary scattering reduces phonon mean free path and reduces thermal conductivity. For instance, the thickness of our Ga$_2$O$_3$ nano-membrane is 427 nm. The phonons with mean free path larger than 427 nm have large possibility to scatter with the film boundaries. The film boundary scattering limits the phonon mean free path in the cross-plane direction and correspondingly reduces cross-plane thermal conductivity. Size effects in nanoscale Ga$_2$O$_3$ electronics would result in further-reduced thermal conductivity, leading to heat dissipation problems in these devices. This highlights the demand of proper thermal management for Ga$_2$O$_3$ electronics.

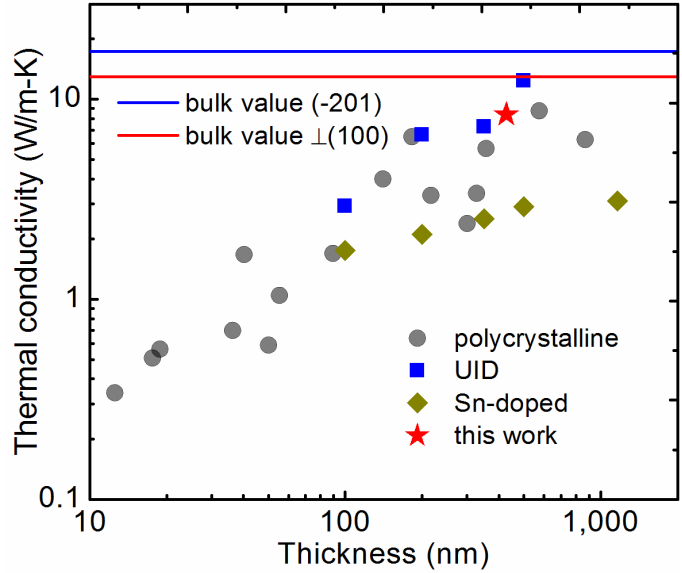

Figure 3. Thickness dependent thermal conductivity of $Ga_2O_3$ thin films. The data for polycrystalline films is from reference.[33] The data for the unintentionally doped (UID) and Sn-doped (-201) orientated thin films is from reference.[34] The blue and red lines are the bulk values in (-201) orientation and perpendicular to (100) orientation.[3, 34]

To understand the phonon transport across perfect $Ga_2O_3$-diamond interface, a Landauer approach[35-38] with transmission function from diffuse mismatch model (DMM) is applied to calculate the TBC at the $Ga_2O_3$-diamond interface. The general form of the Landauer formula is:

$$G = \sum_p \frac{1}{2} \iint D_1(\omega) \frac{df_{BE}}{dT} \hbar\omega v_1(\omega) \tau_{12}(\theta, \omega) \cos\theta \sin\theta \, d\theta d\omega , \qquad (1)$$

where $D$ is the phonon DOS, $f_{BE}$ is the Bose-Einstein distribution function, $\hbar$ is the reduced Planck constant, $\omega$ is the phonon angular frequency, $v$ is the phonon group velocity of material 1 ($Ga_2O_3$), $\tau_{12}$ is the transmission coefficient from material 1 to 2 (here it is from $Ga_2O_3$ to diamond), $\theta$ is the angle of incidence, and the sum is over all incident phonon modes. The expression of the transmission function from DMM[39] is

$$\tau_{12}(\omega) = \frac{\sum_p M_2(\omega)}{\sum_p M_1(\omega) + \sum_p M_2(\omega)}, \qquad (2)$$

where $M$ is the phonon number of modes. Because the transmission function from DMM does not depend on the angle of incidence, the Landauer formula can be simplified as:

$$G = \sum_p \frac{1}{4} \int D_1(\omega) \frac{df_{BE}}{dT} \hbar \omega v_1(\omega) \tau_{12}(\omega) d\omega. \qquad (3)$$

The phonon properties of diamond are obtained from first principles calculation with VASP, and the phonon properties of $Ga_2O_3$ are from Materials Project.[40-42]

The calculated phonon transmission coefficients are shown in Figure 4 (a). The low transmission coefficients at low frequency is derived from the acoustic branches. The number of phonon modes of three dimensional material is proportional to the square of wavenumber, which equals to phonon angular frequency over group velocity for acoustic branches at low frequency, as the phonon dispersion relation is almost linear near the Gamma point. For the longitudinal acoustic (LA) polarization, the phonon group velocity of diamond perpendicular to the (100) plane is 17553 m/s, while the phonon group velocity of $Ga_2O_3$ perpendicular to the (100) plane is 6809 m/s, which means that the group velocity of diamond is 2.58 times that of $Ga_2O_3$, and the number of modes of diamond LA at low frequencies is only 15% of that of $Ga_2O_3$. As a result of the large difference between the acoustic group velocities, the phonon transmission coefficient is very low at low frequencies. For most interfaces involving diamond, the transmission is usually low because of the large phonon group velocity and high cutoff frequency of diamond. At high frequencies, the wavenumber of $Ga_2O_3$ at a certain frequency is relatively close to that of diamond, and the transmission coefficient increases.

The calculated TBC from the Landauer approach is 312 MW/m$^2$-K as shown in Figure 4 (b) at the Ga$_2$O$_3$ cutoff frequency. Because of the complex crystalline structure (large unit cell) of Ga$_2$O$_3$, there are a large number of optical phonon modes. As a result, acoustic phonons of Ga$_2$O$_3$ only contribute to about 8% of the total TBC while optical phonons contribute to about 92%. The calculated TBC is significantly larger than the measured value. The difference between the theoretical TBC and the measured TBC is attributed to the interfacial bonding and real contact area at the interface as discussed above. The van der Waals bonding at Ga$_2$O$_3$-diamond interface is much weaker than covalent bonds, which reduces TBC significantly.[17, 20, 29-30] The calculated results shed light on the possible TBC with perfect interface, which guide the material growth and device design.

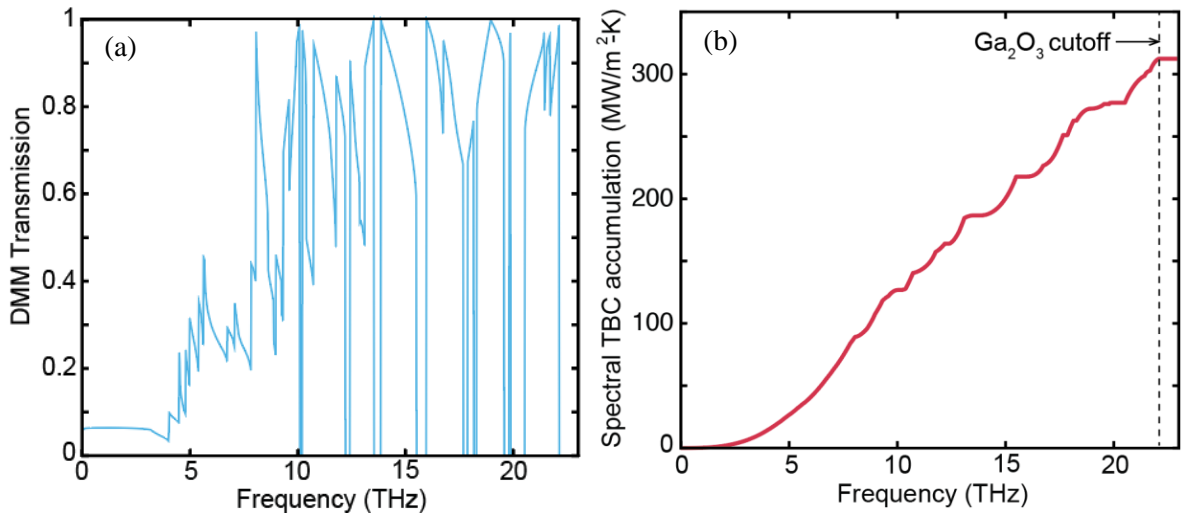

Figure 4. (a) The transmission coefficient from DMM at Ga$_2$O$_3$-diamond interface. (b) The spectral TBC accumulation at Ga$_2$O$_3$-diamond interface from Landauer approach.

To understand the impact of the Ga$_2$O$_3$–substrate TBC on the thermal performance of a power device, we use an analytical solution for the temperature rise in multilayer structures with discrete heat sources.[43] The modeled device consisted of a 500-nm (100) $Ga_2O_3$ layer atop a substrate consisting of either high quality diamond, SiC, or Si. The $Ga_2O_3$ layer was prescribed an anisotropic thermal conductivity with $k_z$ = 12 W/m-K and $k_r$ = 21 W/m-K in accordance with published values.[32] The modeled device structure was a 10 finger device with 50 µm gate-to-gate spacing. The heat sources were each assumed to be 4 x 150 µm and the total domain was 2000 x 2000 µm. A total power density of 10 W/mm was applied to the simulated device. The device structure and simulated heating can be seen in Figure 5 (a-b). Figure 5(a) shows the schematic diagram of the modeled device demonstrating the cross section and heat source spacing. Figure 5(b) shows the top view of the simulated device (a $Ga_2O_3$-diamond TBC of 100 MW/m²K on a diamond substrate with a thermal conductivity of 2000 W/m-K). As expected, the applied power leads to an increase in peak temperature at each $Ga_2O_3$ finger.

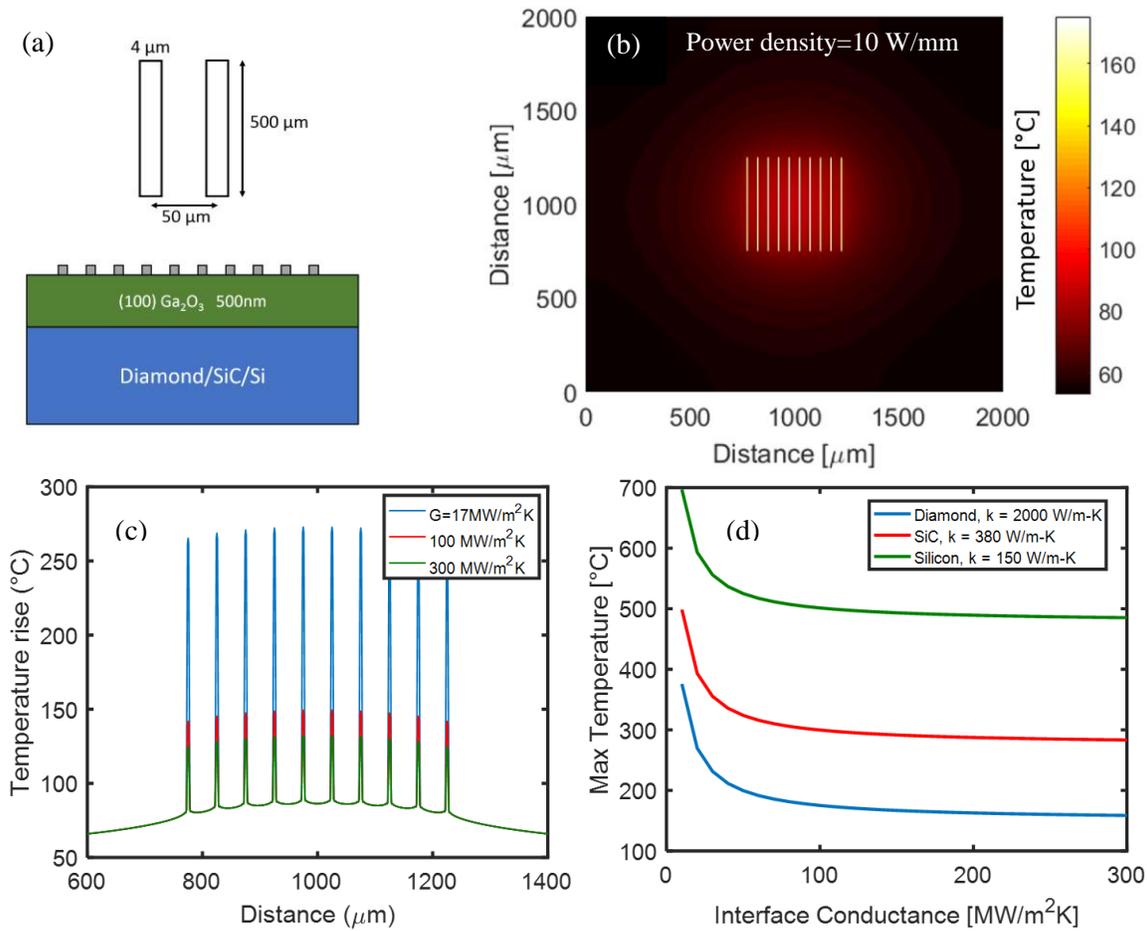

Figure 5. (a) Schematic of the modeled device. (b) Top view temperature field of the simulated device. (c) Heating profile of $Ga_2O_3$ on diamond devices with a TBC value of 17, 100, and 300 MW/m$^2$K, respectively. (d) The maximum temperature of the device for a diamond, SiC, or Si substrate as a function of varying TBC between $Ga_2O_3$ and substrates.

The impact of the $Ga_2O_3$-substrate TBC was evaluated by adjusting its value in the model from 10 to 300 MW/m$^2$K for each substrate material. Figure 5(c) shows the device temperature distribution across the center of the fingers for a simulated $Ga_2O_3$ on diamond device with a TBC value of 17, 100, and 300 MW/m$^2$K, respectively. The decrease in TBC would increase the temperature rise significantly, especially when the TBC is not large. As shown in Figure 5(d), when the $Ga_2O_3$-substrate TBC is small, TBC is the dominant factor limiting heat dissipation. When TBC goes beyond 70 MW/m$^2$-K, substrate thermal conductivity is the dominant factor which limits device thermal dissipation. Additionally, regardless of the TBC value, it is crucial to have a high thermal conductivity substrate. For instance, a SiC substrate with a TBC of 300 MW/m$^2$K shows a maximum temperature of 283 °C, while using a diamond substrate with a TBC of 17 MW/m$^2$K (much lower than 300 MW/m$^2$K), as measured in this work, results in a similar maximum temperature rise. This result demonstrates the importance of implementing a high thermal conductivity substrate such as diamond into $Ga_2O_3$ power devices. Because of the relatively low thermal conductivity of $Ga_2O_3$, even a low TBC value (but equal to or larger than 17 MW/m$^2$K) for a device on diamond will outperform a device on a SiC substrate with an exceptional TBC. This will be useful in guiding device design when integrating $Ga_2O_3$ with high thermal conductivity substrates.

## 3. CONCLUSIONS

A possible solution to cool $Ga_2O_3$ electronics is to integrate thin $Ga_2O_3$ membranes with diamond to fabricate $Ga_2O_3$-on-diamond devices by taking advantage of the ultra-high thermal conductivity of diamond. A good understanding of the TBC between $Ga_2O_3$ and diamond is still lacking. In this work, we measured the TBC of the interfaces of smooth exfoliated $Ga_2O_3$ and polished single crystal diamond. The longitudinal phonon group velocity in the direction perpendicular to the (100) plane of $Ga_2O_3$ is 6620 m/s, which matched very well with DFT-calculated value (6809 m/s). Reduced thermal conductivity of the $Ga_2O_3$ nano-membrane (8.4 ±1.0W/m-K) was observed and attributed to size effects (phonon-boundary scatterings). The Van der Waals $Ga_2O_3$-diamond TBC was measured to be 17 -1.7/+2.0 $MW/m^2$-K, which is in the TBC range of transfer-printed metal films and comparable to the TBC of several physical-vapor-deposited diamond interfaces. This value is relatively quite high by taking the weak bonding strength and small contact area into consideration. The TBC calculated with a Landauer approach and DMM is 312 $MW/m^2$-K, which sheds light on the possible TBC we can achieve. The thermal performance of $Ga_2O_3$-on-diamond devices were modeled to study the effect of $Ga_2O_3$-substrate TBC and substrate thermal conductivity. Our study is important for both applications of power electronics thermal management, and fundamental understanding of heat transport across Van der Waals interfaces.

## ACKNOWLEDGEMENTS

The authors would like to acknowledge the funding support from the Office of Naval Research under a MURI program. Research at NRL was supported by the Office of Naval Research.